\documentclass[]{aa}
\usepackage{times,graphics,rotate,psfig,epsfig,natbib}
\bibpunct{(}{)}{;}{a}{}{,} 
\newcommand{\HI} {H\,{\sc i}}

\newcommand{\NV} {N\,{\sc v}}

\newcommand{\CIV} {C\,{\sc iv}}

\newcommand{\SiIV} {Si\,{\sc iv}}

\def\<<{{\ll}}
\def\>>{{\gg}}
\def\=={{\equiv}}

\def\v0{v_{0}}

\def\beq{\begin{equation}}
\def\eeq{\end{equation}}
\def\beqa{\begin{eqnarray}}
\def\eeqa{\end{eqnarray}}

\begin{document}
\title{An analysis of STIS HST UV spectra of M33 early B supergiants\thanks
{Partly based on INES data from the IUE satellite}}
\author{M.A.~Urbaneja\inst{1}, A.~Herrero\inst{1,2}, R.P.~Kudritzki\inst{3},
F.~Bresolin\inst{3}, L.J.~Corral\inst{1}, J.~Puls\inst{4}}
\offprints{M.A. ~Urbaneja} 
\mail{maup@ll.iac.es}
\institute{Instituto de Astrof\'\i sica de Canarias, E-38200 La Laguna, 
Tenerife, Spain
\and
Departamento de Astrof\'{\i}sica, Universidad de La Laguna,
Avda. Astrof\'{\i}sico Francisco S\'anchez, s/n, E-38071 La Laguna, Spain
\and
Institute for Astronomy, University of Hawaii, 2680 Woodlawn Drive, 
Honolulu, Hawaii 96822, USA
\and
Universit\"ats-Sternwarte M\"unchen, Scheinerstr. 1, D-81679 M\"unchen, 
Germany}

\authorrunning{Urbaneja et al.}
\titlerunning{STIS UV spectroscopy of M33 early B supergiants}

\date{received 07-02-02; accepted 04-03-02}

\abstract{
We present terminal velocities of M33 B-supergiants, obtained
from STIS HST spectra 
as part of our programme to investigate the Wind Momentum --
Luminosity Relationship (WLR) in the Local Group.
Terminal velocities are derived from their \ion{N}{v},
\ion{C}{iv}, and \ion{Si}{iv} resonance lines in UV spectra. Comparing with
IUE spectra of Galactic B-supergiants we found evidence
of low metallicity in three of our objects.
The terminal velocities are consistent with the corresponding values of Galactic
stars, except for B-133. For this
star we find a very large $v_\infty$ and a red \ion{Si}{iv}
component deeper than the blue one, that might be an indication
of binarity. The average ratio
between terminal and turbulent wind velocities is 0.25,
well above the value found for Galactic stars.

}

\maketitle

\section{Introduction}

The Wind Momentum--Luminosity Relationship \citep[WLR, ][ for recent reviews see
Kudritzki 1998 and Kudritzki \& Puls 2000] {kud95,puls96}
offers the very attractive
opportunity to derive the stellar luminosity directly from the analysis 
of the observed spectrum, provided that the wind is radiatively
driven. Accurate individual stellar distances can then be 
obtained if the apparent magnitude and extinction are known. 
This method, that has been extensively 
described elsewhere (see for example the reviews cited above), 
can be potentially applied to individual stars up to distances of 10--20 Mpc, 
reaching the Virgo and Fornax clusters with accuracies as
low as 0.1 mag, using present 
day telescopes and techniques \citep{mccarthy97,kud99}.

However, accurate application of the method requires previous
calibrating work, as the WLR depends on the stellar metallicity
and the spectral type \citep{puls96,kud99}.
Thus observation and analysis of stars of 
different spectral type and metallicity at known distances are 
crucial to calibrate the method before applying it.

For this reason we have started in our group a number of programs aimed at
calibrating the WLR for different stars in different environments within
the Local Group \citep{puls96,mccarthy97,kud99,h01}. 

In the present paper, we analyze STIS HST UV spectra of
M33 B--supergiants. Our analysis is primarily aimed at obtaining the
wind terminal velocities, needed to calculate the WLR, that
will be determined after the analysis of the optical spectrum.

In addition our study constitutes, together with a similar study of
B-supergiants in M31 \citep{bres02}, 
the largest homogeneous sample of stars analyzed in the UV beyond 
the Milky Way and the 
Magellanic Clouds \citep[see][ for a recent list]{Prinja98},
and an important increase in the total number 
of analyzed B-supergiants, in particular those with large intrinsic
brightness (bright magnitudes are of course still a primary selection
criteria for our extragalactic targets). The group of early B--supergiants
is particularly important for the spectral type dependence of the WLR.
Between O9 and B3, two jumps are observed in the WLR, one between 
O9 and B0 and another between B1 and B1.5, while the WLR for B-supergiants
between B1.5 and B3 is not well understood in terms of the radiatively
driven wind theory, as wind momenta lower than expected are derived for the
analyzed Galactic stars \citep{kud99}.

In Sect.~\ref{secobs} we show the observations of the spectra, that are
described in Sect.~\ref{desesp}. Sect.~\ref{anal} presents the analysis
of the UV profiles, while in Sect.~\ref{indiv} we discuss the individual
aspects found for each star. In Sect.~\ref{results}
we present the discussion of the results and the conclusions.

\section{The observations}
\label{secobs}

Observations were made with the HST STIS, using the GL140
grating which provides a resolution from 310 to 210  km s$^{-1}$ in the
wavelength range from 1150 to 1700 \AA. We used a 0.$^{\arcsec}$2
wide slit, which is recommended for optimizing the spectral purity.
Table~\ref{obs} gives our
list of objects and other details of the observations.

\begin{table*}
  \caption[]{M33 stars observed with HST/STIS. The first column gives
the star identification in the catalog of \cite{iv93} (IFM93), while the
second one gives the identification by~\cite{hum80} (HS80).
Coordinates are taken from~\cite{iv93} and precessed for equinox 2000., 
except for B-38 and B-133, for which coordinates from~\cite{mas96} are
taken. Magnitudes are from the same sources, while spectral types are
from~\cite{monteve96} and~\cite{monteve00} , except for 0900 and 1137, for which new spectral
types are given (that will be justified in a forthcoming paper)}

\label{obs}
  \begin{center}
    \begin{tabular}{r r c c r l r l c}
      \hline 
Ident& Ident& $\alpha$(2000)& $\delta$(2000)& $V$& Spectral& Obs.& Exp.& S/N \\
IFM93& HS80 &            &               & mag.& Type   & Date& time(s)&     \\
      \hline 

 0900 &      & 01 33 44.7 & 30 36 18 & 17.3 & B0-B1 I & Jul-8-00 & 5193 & 24 \\
 0785 & 110-A& 01 33 41.0 & 30 22 37 & 16.1 & B1 Ia+  & Aug-7-99& 5201 & 29 \\
      & B-38 & 01 33 00.8 & 30 35 05 & 16.7 & B1 Ia   & Sep-5-00 & 5205 & 30 \\
 0515 & B-133& 01 33 29.0 & 30 47 44 & 17.6 & B1.5 Ia & Aug-5-99 & 5205 & 30 \\
 1733 & B-526& 01 34 15.9 & 30 33 46 & 17.6 & B2.5 I  & Dec-11-00 & 5201& 15 \\
 1137 &      & 01 33 53.1 & 30 35 28 & 16.7 & B3 I    & Jul-9-00 & 5205 & 15 \\
\end{tabular}
\end{center}
\end{table*}

There are two spectra for each star. We merged these two spectra
after checking their relative displacement by cross-correlating them.

Merged stellar spectra are corrected for the relative
velocity between stars and observer. This is done by measuring
the displacement from rest wavelength of metal lines that are purely
photospheric. There are a few photospheric lines that
could in principle be used in the observed spectral range
\citep{Prinja90}. They are difficult to identify in all spectra.
Thus, after inspection of the spectra, 
we decided to use the
\ion{Si}{iii} doublet at $\lambda \lambda$ 1500.24, 1501.19 \AA, because it
is the only one that can be clearly identified in all spectra. 
The spectral resolution does not allow us to resolve
both components of the \ion{Si}{iii} doublet. We use as rest wavelength 
that of a
composed line centered at $\lambda$ 1500.72 \AA.~ Because of the 
small signal to noise ratio of the spectra, we
decided to correct one of the spectra of higher quality and then refer the
others to this one. We selected the spectrum of M33-0900 for that purpose.
Once the M33-0900 spectrum had been shifted so that the \ion{Si}{iii} line 
was at rest wavelentgh, we checked this correction with the position 
of other strong interstellar lines \citep[see f.e. ][]{Prinja90}, as these
are probably produced by the interstellar medium close to M33-0900.
We can see in Fig.~\ref{uvesp} that the shifted interstellar lines
are at their expected rest wavelengths.

To correct the other spectra we cross-correlated them with the corrected
spectrum of M33-0900. To avoid possible biases, strong P-Cygni profiles 
have been masked\footnote{Spectral regions of
\ion{N}{v}, 
\ion{C}{iv}, \ion{Si}{iv} UV resonance lines have been set to zero}
before correlations \citep[see ][]{Howarth97}. 
For M33-1137 and M33-B-526, the stars with lowest SN spectrum, 
the cross-correlation function does not present a well defined maximum.
Checking the strong interstellar lines present in the spectra (\ion{Si}{ii}
$\lambda\lambda$ 1260.40, 1526.70) we found good
concordance between all spectra, excluding M33-1137 and M33-B-526. 
We decided to shift both spectra until the position of their
interstellar lines agree with those of other spectra. 
Special care was taken in the
case of M33-B-526, as the unshifted spectrum shows several lines 
in the vicinity of the rest wavelength of the intestellar lines.

After having shifted the spectra we rectified the continuum by
tracing a polynomial through a number of selected continuum points chosen
iteratively, in the same way as described in \cite{h01}.

\section{Description of the spectra}
\label{desesp}

\begin{figure*}
\centering
{\psfig{figure=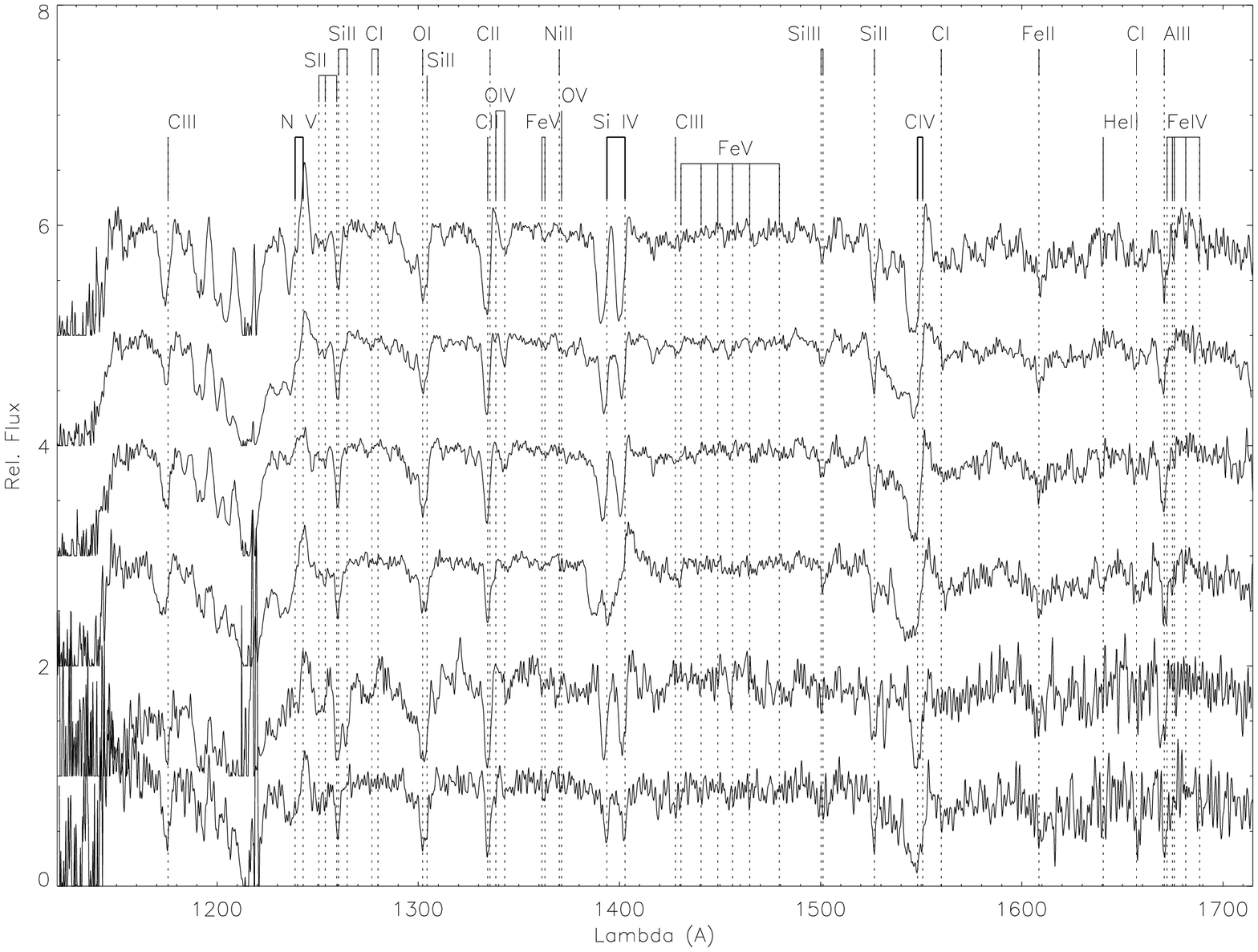,width=15.0cm,height=18.0cm,angle=90}}
\caption[]{The observed rectified UV spectra. From top to bottom: 
M33-0900, M33-110-A, M33-B-38, M33-B-133, M33-B-526
and M33-1137. Relative fluxes have been
arbitrarily displaced in ordinates for the sake of clarity. The main IS
lines are marked at the top, and below we have indicated the rest
wavelengths of the most important stellar lines.}
\label{uvesp}
\end{figure*}

After all the above corrections have been performed, we
finally obtained the spectra displayed in Fig.~\ref{uvesp}.

\begin{figure*}
\centering
{\psfig{figure=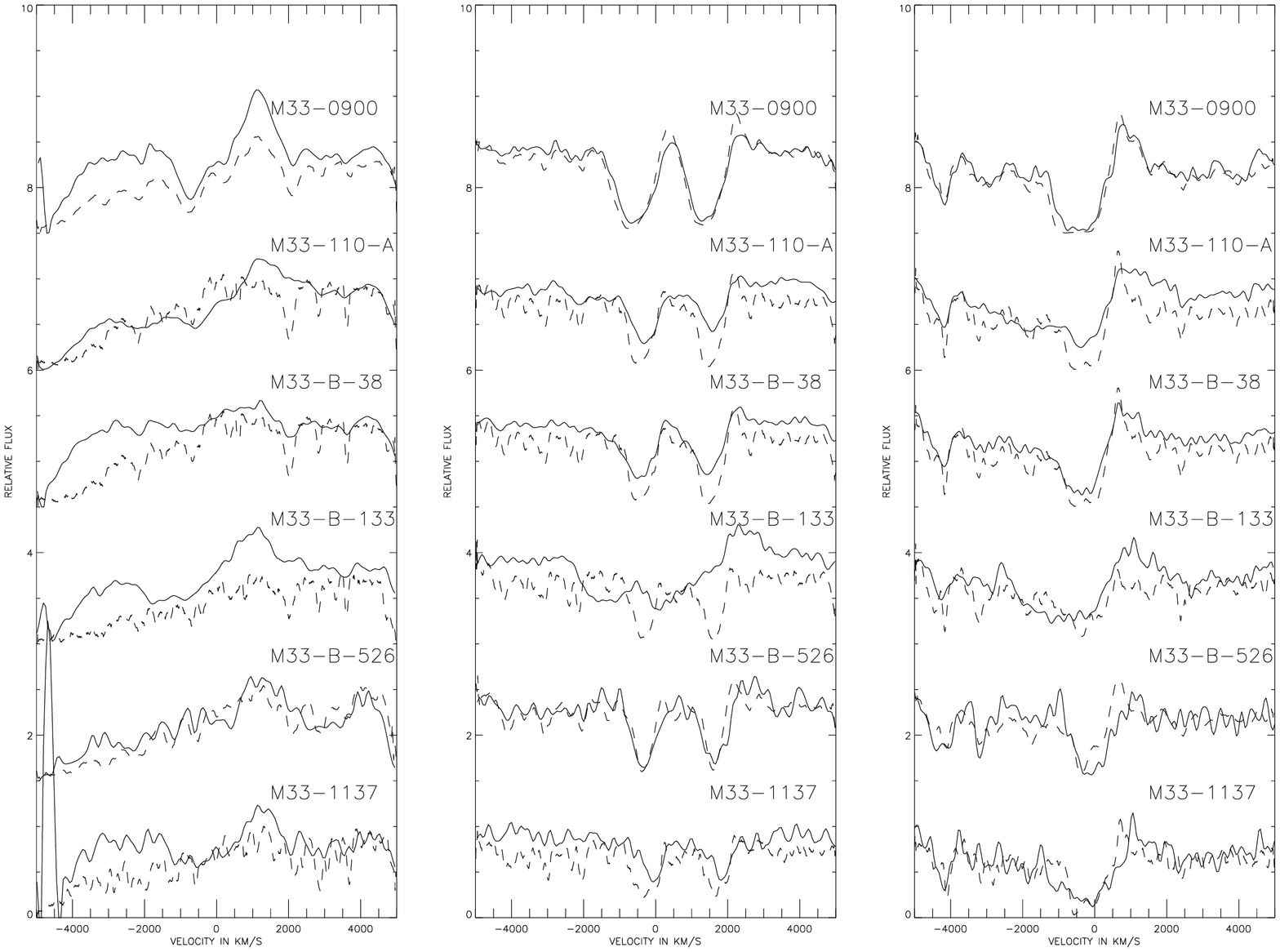,width=15.0cm,height=16.0cm,angle=90}}
\caption[]{Comparison of the \ion{N}{v} (left), \ion{Si}{iv} (middle) and \ion{C}{iv} (right)
lines in the M33 (full line) and Galactic (dashed line) stars. From top to bottom: 
M33-0900 and HD\,154090; M33-110-A and HD\,148688; M33-B-38 and
HD\,148688; M33-B-133 and HD\,38711; M33-B-526 and HD\,198487;
M33-1137 and HD\,51309. Spectral clssifications are given in
Table~\ref{params}. Relative fluxes have been
arbitrarily displaced in ordinates for the sake of clarity. Galactic spectra
have been degraded to the M33 HST resolution.}
\label{m33mw}
\end{figure*}

One of the difficulties of stellar analyses in M33 is the large possible
variation in metallicity, that complicates the interpretation of the
spectrum. Thus in Fig.~\ref{m33mw} we show a comparison of
the \ion{N}{v}, \ion{Si}{iv} and \ion{C}{iv} lines of the M33 stars with
Galactic stars of similar spectral classification. Their identifications
and spectral types are given in Table~\ref{params}.

The Galactic UV spectra have been taken
from the INES IUE archive, and have been degraded from the
original high resolution to the lower resolution of our M33 HST spectra.
We emphasize that we have taken the rectification points of the
M33 stars as a guide to rectify the IUE spectra in the same way.
Therefore, the extra absorption seen redwards of the \ion{Si}{iv}
and \ion{C}{iv} profiles in some Galactic stars are probably
an effect of metallicity. Furthermore these extra
absorptions are consistent with the profiles indicating a lower
metallicity in the M33 stars. The \ion{N}{v} comparison is strongly
affected by the different L$\alpha$ absorptions. 

{\bf M33-0900} has \ion{Si}{iv} and \ion{C}{iv} profiles
that are very similar to those of HD 154090. Although
\ion{N}{v} is stronger in M33-0900, the star looks
normal for its spectral type.

\begin{figure}
\centering
{\psfig{figure=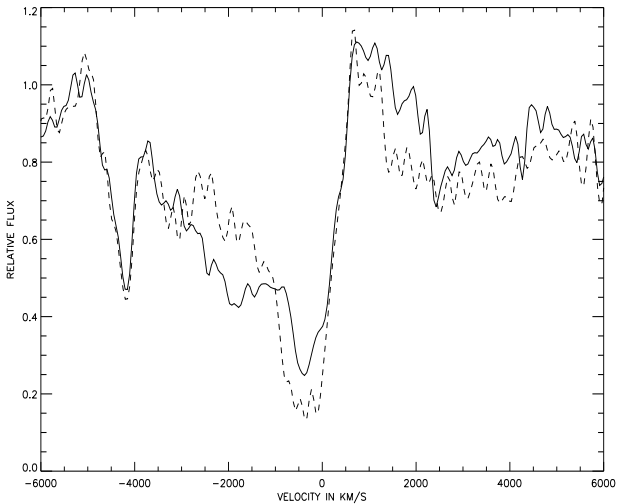,width=7.0cm,height=8.5cm,angle=90}}
\caption[]{Comparison of the \ion{C}{iv}
line in M33-110-A (solid line) and M33-B-38 (dashed line)}
\label{110ab38}
\end{figure}

This is not the case for {\bf M33-110-A},
which in comparison to HD 148688 displays very weak Si and C lines. Moreover
considering the luminosity effect we would expect for 
\ion{Si}{iv}, the difference is even larger. This is consistent with
the low Si and O abundances derived by \cite{monteve00}. Thus we can expect
a low metallicity for 110-A, also consistent with the higher
continuum level redwards from \ion{Si}{iv} and \ion{C}{iv}.
The \ion{N}{v} profile is again stronger than in
the comparison star. The slopes
of the blue wings of both the Si and C lines are smaller than in
HD\,148688. This will be interpreted during the
fit procedure as a larger dispersion velocity. Finally we shall mention
the broad absorption in the \ion{C}{iv} blue wing that becomes more 
apparent due to the low absorption in this line. This additional
absorption is present in both individual M33-110-A spectra and can be
clearly indentified in Fig.~\ref{110ab38} where we show a comparison of the
\ion{C}{iv} line in 110-A and in B-38. We have used this figure as
a guide in fitting the \ion{C}{iv} line in 110-A.

{\bf M33-B-38}, for which \cite{monteve00}
again derive low O and Si abundances, has also been compared with HD 148688.
Similar comments as for 110-A
are applicable, although effects are less extreme
and the \ion{N}{v} profile in this case is marginal (as in the Galactic star).

This has previously been observed and 
analized by \citet{Bianchi96} and was later reanalyzed by \citet{Prinja98}. 
Both used the same HST GHRS spectrum, which looks like ours,
including the lower slope in the blue wings as compared to Galactic
stars \citep[see Fig. 4 from ][]{Bianchi96}.
\citet{Bianchi96} find a narrow absorption feature (narrow absorption
component, NAC) at a 
blueshifted velocity of $\sim$1250 km s$^{-1}$, 
which is close to the edge velocity determined by 
\cite{Prinja98}, 1225 km s$^{-1}$. Interestingly, we also find evidence of
a NAC at the same velocity, which may indicate that this is a more
permanent feature.

{\bf M33-B-133} shows the most puzzling spectra. The red component
of the \ion{Si}{iv} doublet is deeper than the blue one, which cannot
be explained by the theory. The effect is
due neither to noise nor to variability, as it is present
in the same form in both individual spectra of M33-B-133. 
We see numerous features
in the \ion{C}{iv} profile that could be interpreted as NACs.
From them, only the absorptions at 250, 585 and 1990
km s$^{-1}$ show both components. 
None of these NACs, however, seems strong enough
to explain the red component of \ion{Si}{iv}. 

This is not the only remarkable feature in the spectrum of M33-B-133.
This object has been classified by
\citet{monteve00} as B1.5 Ia, and is compared with HD\,152236
(B1.5 Ia+)
in Fig.~\ref{m33mw}. We see that the profiles are quite different.
In fact, the spectrum of B-133 resembles more that of
$\kappa$ Ori (HD\,38771), a B0.5 Ia star, although a comparison
of the optical spectra excludes such an early spectral type for
M33-B-133. 

Another possibility is a blend with the spectrum of a hotter star.
This is supported by the resemblance of the spectra of B-133 and
$\kappa$ Ori, and by the data in Table~\ref{obs}. Note that there
B-133 has a larger S/N than would be expected from a comparison
of its visual magnitude and spectral type with those of the
other stars. Thus, although a particular set of circumstances
is required because the B1.5 supergiant dominates the blue
spectrum, it is not impossible that a hotter companion of type
O9.5-B0.5 is dominating the UV spectrum. If this is the case
the apparent strong 
red component of the \ion{Si}{iv} doublet would be the 
main contribution of the B1.5 star in the UV spectrum.

However, we could not reproduce the observed spectrum 
by combining IUE spectra of different spectral types
without extra {\em ad hoc} assumptions (as arbitrary 
relative displacements), nor
can we detect any clear indication of binarity in the
HST WFPC2 images of B-133, or other features indicative
of a companion in the blue and UV spectrum, and thus the suspected
binary nature of B-133 remains unconfirmed.

{\bf M33-B-526} displays lines very similar to 
those of HD 198487, in spite of its suspected binary nature
from its appearance in the Keck I screen (McCarthy, private communication).

The \ion{C}{iv} profile displays several
absorption features redwards of the rest wavelength that could be
interpreted as contributions from a companion star. However, their
positions do not correspond to those of the doublet separation
Thus we expect this to be the spectrum of a single star.

{\bf M33-1137} has \ion{Si}{iv} profiles that are clearly weaker than those
of the selected Galactic counterpart, HD\,51309. The \ion{C}{iv} profile
in M33-1137 is clearly broader than in HD\,51309, with a shallow
slope in the blue wing. It displays additional absorptions that agree
with the positions of the \ion{C}{iv} components at $-$75 km s$^{-1}$,
thus clearly pointing to wind inhomogeneities. Weaker absorptions are
compatible with components at $-$550 km s$^{-1}$. Other absorptions
cannot be clearly assigned to the \ion{C}{iv} doublet, specially
considering the low SNR. The \ion{N}{v} feature is more evident in M33-1137
than in the Galactic star. 
The observed spectrum points again to a lower metallicity
in the M33 star than in the Galactic comparison star.

Therefore the UV spectral morphology of the observed M33 stars separates
them in three groups: a group (particularly evident in \ion{Si}{iv}) with 
lines  weaker than
their Galactic comparison stars (M33-110-A, M33-B-38 and M33-1137),
a second group with lines similar to their Galactic counterparts
(M33-0900 and M33-B-526) and one
star (M33-B-133) showing peculiar profiles for its spectral type.
In addition, the M33 stars
seem to have lower slopes in the blue wings, pointing to larger
dispersion velocities. This is particularly clear in the first
group of stars.

\section{Analysis of the resonance lines}
\label{anal}

We use the method described by Haser \citep[][ see also Lamers et al. 1999]{has95} to analyse the UV P Cygni lines in order to derive wind
terminal velocities. We have also derived \HI\, column densities towards 
the M33 stars by fitting the IS L$\alpha$ line, in the same way as 
in \citet{h01} \citep[see also ][]{jen70,boh75}.
The values quoted in Table~\ref{params} for these column densities
are similar to those obtained for stars in our
Galaxy \cite[see Fig.2 of][]{shu85}.

After continuum
rectification, we extract small regions of each spectrum centered
in our strategics doublets: \ion{N}{v} $\lambda\lambda$ 1238.819,
1242.798; \ion{Si}{iv} $\lambda\lambda$ 1393.73, 1402.73 and \ion{C}{iv}
$\lambda\lambda$ 1548.191, 1550.761. In case of
discrepancies or difficulties, we give more weight to the
\ion{Si}{iv} and \ion{C}{iv} lines than to the \ion{N}{v} one.

To get the wind terminal velocities
we have to correct for the underlying photospheric components,
which we do in an approximate way, by using IUE spectra of hot stars with
weak winds (and projected rotational velocities as low as possible) as
templates. These templates have been convolved with appropriate
rotational profiles, to account for the individual stellar rotational
speeds. We excluded \ion{N}{v} from this procedure, as the corresponding
spectra are noisy and do not show any features. 

We selected a sample of Milky Way dwarfs with spectral types from O9V to B3V
taken from the INES database. This covers 
the spectral type range
of our M33 stars. Selection of photospheric template and continuum
rectification has a big impact when fitting emission peaks,
but has little effect on the determined terminal velocities.
The IUE spectra have been corrected for the relative velocity between stars
and observer, in the same way as for M33 stars.
The stars used for the photospheric templates are specified
in Table~\ref{params}. Note that the profiles of interest are very weak
in these stars, but even so the different metallicity may
be a small source of error.

The velocity stratification is parameterized as a usual $\beta$-law, 
and we account for the wind turbulent velocity in the same manner
as described in \cite{has95} and \cite{h01}.
The indetermination in the exponent of the velocity field, $\beta$,
produces an additional uncertainty in the terminal velocities.

\begin{figure*}
\centering
{\psfig{figure=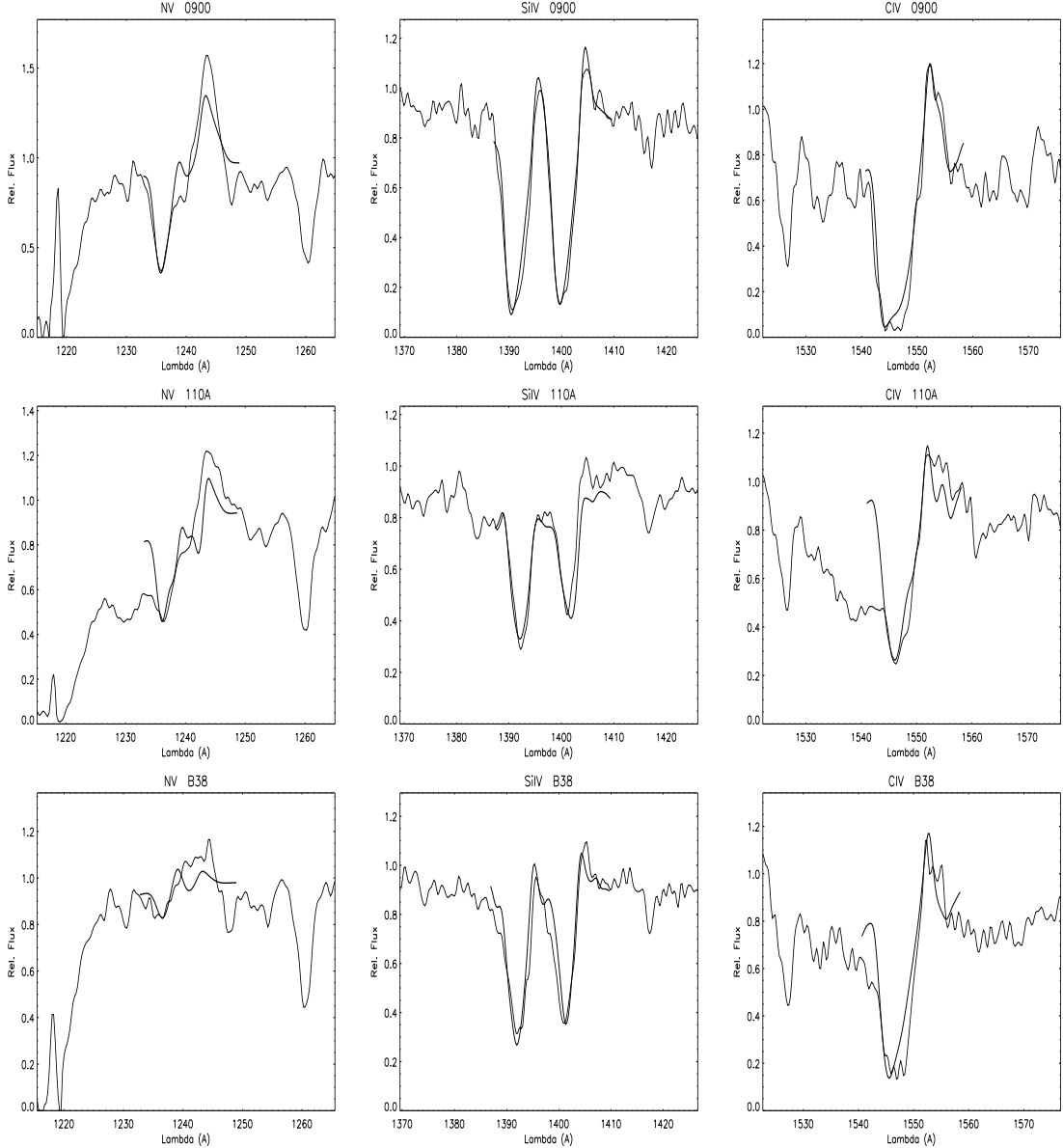,width=16.0cm,height=15.0cm,angle=90}}
\caption[]{Final fits for the three earliest stars of our sample.
Lines are plotted from left to right (\NV, \SiIV, \CIV) and stars from top
to bottom in the same order as they are listed in the tables.}
\label{fitsh}
\end{figure*}

\begin{figure*}
\centering
{\psfig{figure=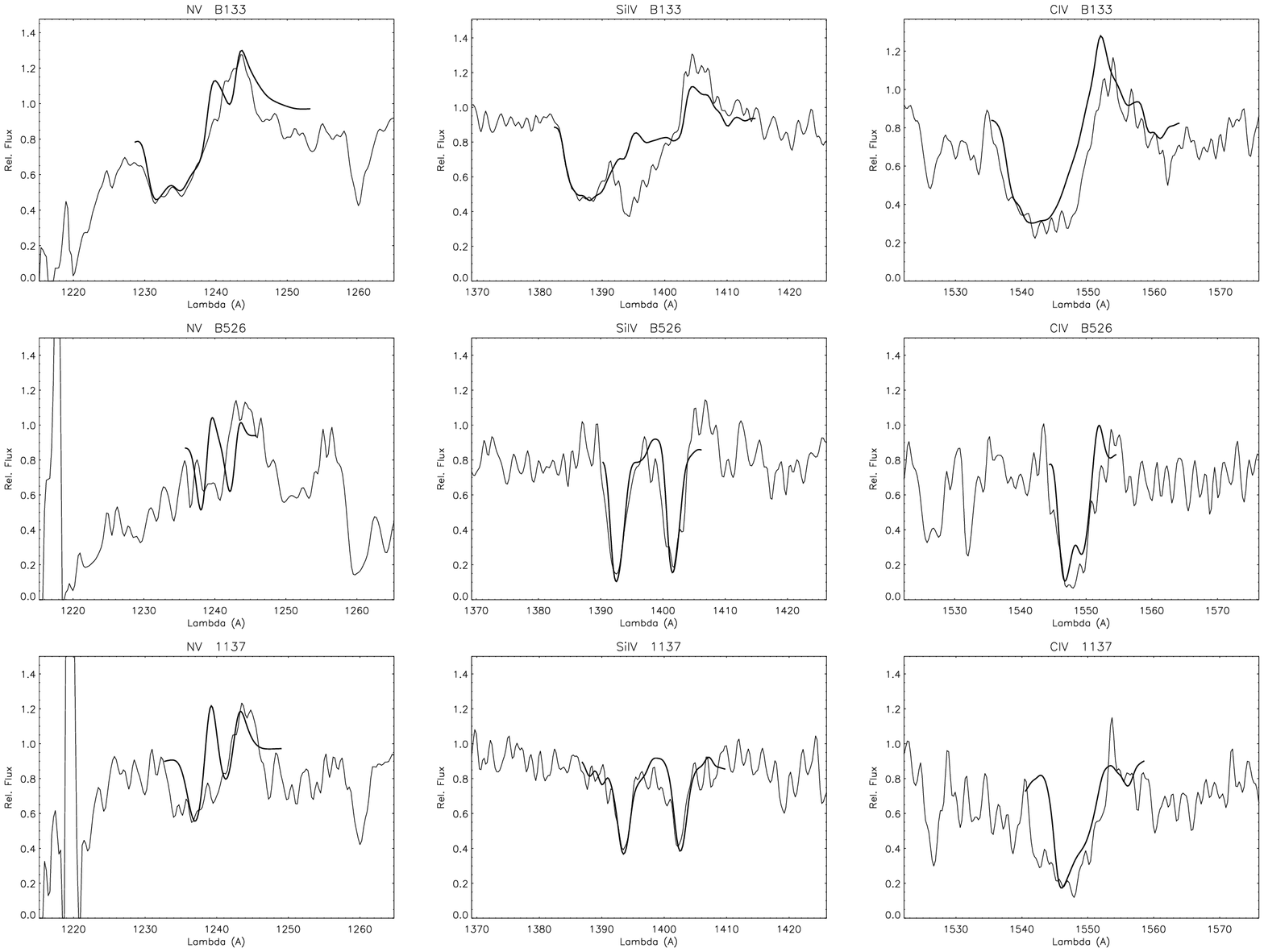,width=16.0cm,height=15.0cm,angle=90}}
\caption[]{As Fig.~\ref{fitsh}, however for the three last stars in 
Table~\ref{obs}}
\label{fitsc}
\end{figure*}

Table~\ref{params} gives the fit
results with respect to the wind and turbulent velocities. 
The final fit to each line is
shown in Figs.~\ref{fitsh} and \ref{fitsc}. 

\section{Individual comments}
\label{indiv}

We briefly comment here on individual aspects of the analysis that could
be of interest.

\subsection{M33-0900}
We used HD\,22591 (B0.5 V) for the photospheric profiles.
The \ion{N}{v} profile shows a large emission peak that we 
cannot reproduce completely (see Sect.~\ref{desesp}).
The $v_{ta}/v_\infty$ ratio is 0.18, larger than in Galactic stars.

\subsection{M33-110-A}
We used HD\,39777 (B1.5 V) for the underlying photospheric
profiles.
We find the same problems as for M33-0900:
a large red emission peak in \ion{N}{v}, and a large
ratio $v_{ta}/v_\infty$, which with a value of 0.25
is much larger than in Galactic B--supergiants.
Furthermore, 
the additional absorption contaminating the blue side of \ion{C}{iv}
makes the fit a bit more difficult, affecting mainly the determination
of $v_{ta}$. This does not seriuosly affect
the uncertainty in the adopted values for $\beta$, $v_\infty$ and 
$v_{ta}$. 

\subsection{M33-B-38}

The terminal velocity of this star has also been determined by 
\cite{Bianchi96} and \cite{Prinja98}. The first authors use an
analysis method similar to the one employed here, and thus it is
not surprising that our value agrees with their. \cite{Prinja98} determine
the terminal velocities of their stellar sample from the violet edges 
of the profiles and the NACs. The velocity they obtain for 
this star is much larger (1225 km s$^{-1}$) than ours
(730 km s$^{-1}$) or the one by Bianchi et al. (about 700 km s$^{-1}$).
Our velocity would support the interpretation that what we have
seen at 1250 km s$^{-1}$ is actually not a NAC.

~\cite{Bianchi96} do not give the turbulent velocity of their stars.
We derive a very large value, 250 km s$^{-1}$, nearly 35$\%$ of
$v_\infty$. 

We adopted a large $\beta$-value, both to improve the
consistency between the \ion{Si}{iv} and \ion{C}{iv} fits, and to
improve the fit of the blue side of the red emission peak.

\subsection{M33-B-133}

We have used HD39777 for phostospheric profiles, as in the case of M33-110-A 
and M33-B-38.

In spite of the low O and Si abundances derived by \citet{monteve00}
for this star, its UV spectrum shows comparatively strong \ion{C}{iv}
and \ion{Si}{iv} P-Cygni profiles, indicating a strong and fast
wind (see Sect.~\ref{desesp}). 
The terminal velocity reaches 2050 km s$^{-1}$. The turbulent
velocity is 150 km s$^{-1}$, a modest 7$\%$. This is the only star for which
we obtain a terminal velocity clearly above the galactic average
for its spectral type \citep[taken from ][]{kp00},
thus challenging the low abundances or
the spectral classification (or both!). 
The spectral type, however, should change from B1.5 to O9 in
order to have a terminal velocity close to the spectral
type average. This is completely ruled out from inspection of the optical
spectrum. 

The fits displayed in Fig.~\ref{fitsc} also show a behaviour
different from those of the other stars. \ion{N}{v} is well fitted. 
The fit of the
\ion{C}{iv} profile is good in the bluest part of the wind
absorption profile but is bad in the rest of the profile.
Fortunately, the first one is the important part for determining
the terminal velocity. The unsatisfactory fit at low
wind velocities is cosmetically very dependent on the underlying
photospheric profile, and thus does not mean very much by itself,
but it would be consistent with the presence of additional
\ion{C}{iv} absorption at low wind velocities.
We keep the low $\beta$ value
as there is no need to increase $\beta$ to improve the consistency
between \ion{C}{iv} and \ion{Si}{iv}.
However, the anomaly in \ion{Si}{iv} indicated in the
preceding section, i.e., a stronger red component, can also
not be fitted and indicates an extra absorption at low 
wind velocities. This uncertainty does not affect the determination
of the terminal and turbulent wind velocities, that have an
accuracy of $\pm$100 km s$^{-1}$.

This does not solve the problem
referred to above. The star has a large $v_\infty$ for its
spectral classification and a \ion{Si}{iv} red component stronger
than the blue one. The first might be attributed to a case of
bi-stability \citep{papu90}, similar to that of P-Cygni in our Galaxy,
where the ratio of terminal velocity to escape velocity is larger.
This however does not explain the red \ion{Si}{iv} component.
One possible solution to this puzzle is that we are looking at
a composite spectrum, B-133 being the star contributing to the
red component of \ion{Si}{iv} (see Sect.~\ref{desesp}). 
If this is the case, we estimate its terminal wind velocity to be of the
order of 450 km s$^{-1}$, but would expect to see extra absorption
in \ion{Si}{iv} at low wind velocities.

\subsection{M33-B-526}

It has been classified as B2.5I \citep{monteve96}, 
from its optical spectrum. We used a B2.5V galactic star
for the photospheric profiles (HD 44402, Z CMa). 

The derived terminal velocity is the
lowest in our sample and the UV spectrum shows only weak signs of mass loss.
However, the turbulence velocity is relatively high, with a
ratio $v_{ta}/v_\infty$ in excess of 0.3. This large value
is the result of a compromise between the fit of \ion{C}{iv}
and that of \ion{Si}{iv}. The first would allow a lower
turbulence, but then we cannot fit the second one. 
Again, the value of $\beta$ has to be large to favour
consistency between both doublets and a better fit to
\ion{C}{iv}. However,
the terminal velocity is not seriously affected. The uncertainty
of both velocities is different now, being that of $v_{ta}$
$\pm$100 km s$^{-1}$ and that of $v_\infty$ $\pm$50 km s$^{-1}$.
For the sum of both we have adopted $\pm$75 km s$^{-1}$.

\subsection{M33-1137}
This is the coolest star of our sample, with spectral type B3Ia.
We take HD32630 (B3V) for the photospheric profiles.

The \ion{Si}{iv} doublet is mainly photospheric or in any case the wind
has a small contribution. The fit to \ion{N}{v} is poor.

The \ion{C}{iv} is the main profile for deriving the parameters
of the velocity field. The low SNR of the spectra and the additional
absorptions described in the preceding section make the fit difficult.
We find the best fit at 750 km s$^{-1}$, 250 km s$^{-1}$ and 2.0
for $v_\infty$, $v_{ta}$ and $\beta$. Again, $v_\infty$ is larger than 
the Galactic average from \cite{kp00}, but now much more
moderately than for B-133.
The uncertainties are slightly larger
than those of the other fits, as already expected from the described
difficulties. We adopt $\pm$100 km s$^{-1}$ for $v_\infty$, $v_{ta}$
and its sum.

\begin{table*}
\begin{center}
  \caption[]{Results obtained for the observed stars. $v_\infty$ and
$v_{ta}$ (turbulence velocity in outer wind) and the uncertainty of
their sum, $\Delta v$, are given in km s$^{-1}$}
\label{params}
    \begin{tabular}{r l l l c c c r c}
      \hline 
Ident& Spectral& Galactic & Spectral& log & $v_\infty$ & $v_{ta}$ & $\Delta v$ & $\beta$ \\
     &  Type   & Star     &  Type   & N(\ion{H}{i}) &  &       & & \\
      \hline 

 M33-0900   & B0-B1 I & HD\,154090 & B0.7 Ia & 20.85 & 950 & 170 & 50 & 1.0 \\
 M33-110-A  & B1 Ia+  & HD\,148688 & B1 Ia  & 21.15 & 800 & 200 & 50 & 1.0 \\
 M33-B-38   & B1 Ia   & HD\,148688 & B1 Ia & 20.70 & 730 & 250 & 50 & 2.0 \\
 M33-B-133  & B1.5 Ia & HD\,152236 & B1.5 Ia+ & 20.97 & 2050 & 150 & 100 & 1.0 \\
 M33-B-526  & B2.5 I  & HD\,198487 & B2.5Ia & 21.10 & 380 & 120 & 75 & 2.0 \\
 M33-B-1137 & B3 Ia   & HD\,51309 & B3Ib & 20.85 & 750 & 250 & 100 & 2.0 \\
 \end{tabular}
 \end{center}
\end{table*}

\section{Discussion and conclusions}
\label{results}

\begin{figure}
\centering
{\psfig{figure=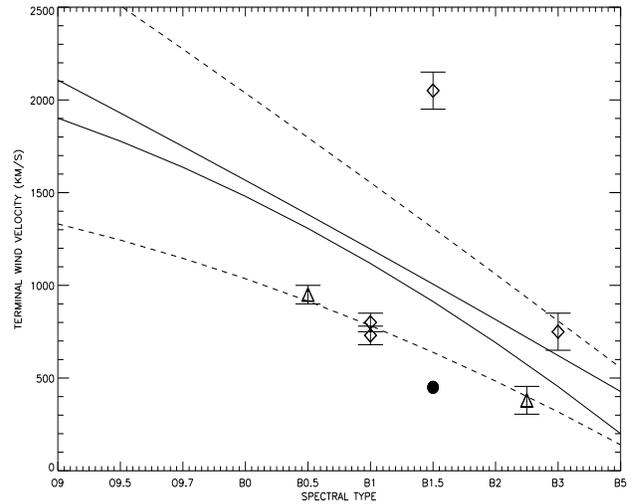,width=7.0cm,height=8.5cm,angle=90}}
\caption[]{The $v_\infty$ obtained for the M33 stars. Solid lines
represent the fits to OI + BIa (lower curve) and
to OII + BIb supergiants (upper curve) of the \citet{kp00} Galactic data. Dashed lines are
$\pm$30$\%$ variations of these curves. The solid dot indicates
the approximate position of B-133 if derive it from fitting 
the red \ion{Si}{iv} component (see text for details)}
\label{vinf}
\end{figure}

In Fig.~\ref{vinf} we have plotted the derived $v_\infty$
against the stellar spectral types, together with parabolic
fits to the average values quoted by \citet{kp00} for
Galactic OB supergiants. The fits have been obtained by
joining OI and BIa supergiants on the one
hand and OII and BIb supergiants on the other.
We have also plotted lines that
indicate a 30$\%$ variation from the plotted average relations, 
a usual scatter range \citep[see Fig.4 in ][]{kp00}.

All M33 B-supergiants
give us values that can be considered normal, except B-133.
We see no difference between the stars with 
different suspected metallicities. This is in agreement with \citet{puls00} 
~\citep[see also ][]{vink00} ~who have argued that the
terminal velocity (which depends on the slope of the line-strength distribution
function, $\alpha$) is primarily controlled by the ratio of light ions vs. iron
group elements, because of the different line statistics. As long as this ratio
is roughly similar, the theoretical expected change of terminal velocity
(due to the "indirect" $\alpha$ effect, see Puls et al., 2000, setc
5.2) is much smaller than if this ratio would be changed. In particular if a
dense wind is present (as is the case for our supergiants) and this ratio 
remains unchanged (which we have to assume for the moment), the effect is
expected to be very
small \citep[see][Fig. 27]{puls00}, since the effective $\alpha$ then remains roughly 
constant. Only for thin winds and/or a significantly lower (general) 
metallicity the effect should become observable.

We could not find a satisfactory explanation for B-133. While
its terminal velocity is well above the average for its spectral
type, it could still be accepted \citep[even with the low metallicity
derived by ][]{monteve00} assuming for example a bi-stability
phenomenon, were it not for the anomalous
red component of the \ion{Si}{iv} doublet. Assuming on the
other hand that B-133 is actually producing the red \ion{Si}{iv}
component, its velocity would be much closer to the average
of its spectral type. However, we could not reproduce the
observed profiles without {\em ad hoc} hypothesis, nor find
conclusive evidence of a binary nature.

Although three of our stars have been analyzed by \citet{monteve00}
we will postpone a discussion of the terminal velocities in terms
of the stellar parameters, as new analyses of their optical
spectra, now including mass-loss effects, are currently under 
way in our group together with a set of newly observed stars.

The wind turbulent velocities derived here are larger than the typical
10$\%$ $v_\infty$ \citep{kp00} or the 14$\%$ found by \citet{h01},
reaching nearly 35$\%$ in the most extreme case. 
Including B-133 we obtain 0.25 for the mean $v_{ta}/v_\infty$
ratio. This result is of the same order of that found by
\cite{bres02} in their analysis of M31 B supergiants.
We tentatively attribute it for the moment
to the extreme character of
our objects, selected among the brightest supergiants in M33,
and as a consequence are among the most luminous objects.

Finally, we have detected evidence of numerous NACs, confirming
the wide presence of wind inhomogenities.

\acknowledgements{
We would like to thank F. Najarro for very useful discussions during
the stellar analyses.
A.H. wants to acknowledge support for this work by the spanish 
DGI under proyect AYA2001-0436, the DGES under
project PB97-1438-C02-01 and from the Gobierno Auton\'omico de
Canarias under project PI1999/008. F.B. and J.P. acknowledge support from the
German DLR, under grant RD-RX 50OR9909/2.}

\end{document}